\documentclass[twocolumn,showpacs,preprintnumbers,amsmath,amssymb,prb,aps]{revtex4-1}
\usepackage{epsfig}
\usepackage{epstopdf}
\usepackage{graphicx}
\usepackage{dcolumn}
\usepackage{bm}
\usepackage{slashbox}
\usepackage{textcomp}
\usepackage{array}
\usepackage{subcaption}
\usepackage{booktabs}

\begin{document}

\title{Density functional study of thermodynamic properties, thermal expansion and lattice thermal conductivity of Fe$_{2}$VAl at high temperature region}

\author{Shamim Sk$^{1,}$}
\altaffiliation{Electronic mail: shamimsk20@gmail.com}
\author{Sudhir K. Pandey$^{2,}$}
\altaffiliation{Electronic mail: sudhir@iitmandi.ac.in}
\affiliation{$^{1}$School of Basic Sciences, Indian Institute of Technology Mandi, Kamand - 175075, India}
\affiliation{$^{2}$School of Engineering, Indian Institute of Technology Mandi, Kamand - 175075, India}


\begin{abstract}
Here, we present the phonon calculations for thermodynamic properties, thermal expansion and lattice thermal conductivity of Fe$_{2}$VAl in the temperature range of $300-800$ K and compared with existing experiment. Phonon dispersion is computed using finite displacement method and supercell approach. The positive frequencies of all the phonon modes indicate the mechanical stability of the compound. The specific heat at constant volume and Helmholtz free energy are calculated under harmonic approximation, while calculation of thermal expansion is done under quasi-harmonic approximation. Lattice thermal conductivity ($\kappa_{L}$) is calculated using first-principle anharmonic lattice dynamics calculations. The zero-point energy and Debye temperature are computed as $\sim$21 kJ/mol and 638 K, respectively. The calculated thermal expansions are found to be $\sim$6.3 $\times$ 10$^{-6}$ K$^{-1}$ and $\sim$7.2 $\times$ 10$^{-6}$ K$^{-1}$ at 300 and 800 K, respectively. A significant deviation between calculated ($\sim$48.6 W/m-K) and experimental ($\sim$22.8 W/m-K) values of $\kappa_{L}$ are observed at 300 K. But, as the temperature increases, the calculated and experimental $\kappa_{L}$ come closer with the corresponding values of $\sim$18.2 W/m-K and $\sim$11.0 W/m-K at 800. The possible reasons for the deviation of $\kappa_{L}$ are addressed. The temperature dependent of phonon lifetime is computed in order to understand the feature of $\kappa_{L}$. Present study suggests that DFT based phononic calculations provide reasonably good explanations of available experimental phonon related properties of Fe$_{2}$VAl in the high temperature range of $300-800$ K.

\vspace{0.3cm}

\end{abstract}

\maketitle
\section{INTRODUCTION}
Phonons play an important role in understanding many properties of the materials, such as thermal conductivity\cite{kaviany,lindsay,kang}, thermal expansion\cite{shastri_4th,shastri_5th,shamim_epl}, thermoelectricity\cite{snyder,liao}, superconductivity\cite{giustino,chen_2018}, mechanical properties\cite{kanchana,petrova} etc. Understanding the phononic transport properties have always been a challenging task at the level of theory as well as computation. This is due to the involvement of many-body interactions, such as electron-phonon interaction, phonon-phonon interaction, phonon-defect interaction etc\cite{ashcroft}. The study of these transport properties becomes more challenging at finite temperature due to the presence of anharmonic effect. However, recent advancement of computational power has enabled us to calculate the phonon related transport properties of the materials at finite temperature that are consistent with the experimental measurement up to some extent. For instance, in the context of thermoelectricity, phonon calculations give two important quantities, \textit{viz.} thermal expansion and lattice thermal conductivity ($\kappa_{L}$), which are the useful quantities to fulfill the evaluation of thermoelectric (TE) materials. Specifically, thermal expansion is the required parameter to select the TE materials in making TE generator. The study of $\kappa_{L}$ is useful to evaluate the efficiency of TE materials. 

TE materials can convert the waste heat into useful electricity\cite{te1,te2}. The performance of TE materials is evaluated by unitless parameter, \textit{figure-of-merit} \textit{ZT}\cite{zt}, defined as $ZT=S^{2}\sigma T/\kappa$. Where, \textit{S} is the Seebeck coefficient, $\sigma$ the electrical conductivity and \textit{T} the absolute temperature of the sample. $\kappa$ is the total thermal conductivity, which consists of the electronic part ($\kappa_{e}$) and lattice part ($\kappa_{L}$). The value of high \textit{ZT} indicates the higher efficiency of TE materials. Efficient TE materials should possess the \textit{ZT} value higher than unity\cite{snyder}. Therefore, the value of \textit{ZT} can be improved by enhancing the power factor ($S^{2}\sigma$) or by reducing the $\kappa$. But, getting high \textit{ZT} is really a challenging task, as \textit{S}, $\sigma$ and   $\kappa_{e}$ are strongly correlated to each other through charge carriers\cite{ashcroft,shamim_mrx}. Hence, optimization of $\kappa_{L}$ is one of the brilliant ways to improve ZT. Therefore, phonon calculation is required in order to optimize $\kappa_{L}$.

For studying the lattice dynamics and phonon related transport properties at finite temperature, an ab initio molecular dynamics simulations have been used in recent decades\cite{molecular1,molecular2,molecular3,molecular4}. But, these techniques demand heavy computational cost. In this regard, the density functional theory (DFT)\cite{dft} based methods come into picture with relatively lower computational cost and reliability. In addition to the electronic structure calculations, DFT also can calculate the phonon dispersion. The displacement of an atom from its original position gives the change in force exerted on each atom. The systematic displacement of atoms gives the phonon frequencies and the method for analyzing these frequencies is called finite displacement method (FDM)\cite{fdm}. Another technique to study the phonon frequencies is density functional perturbation theory (DFPT)\cite{dfpt}. At the DFT level, it is a routine job to use the ground state phonon dispersion to calculate the high temperature phonon related properties. We said that in the context of thermoelectricity, thermal expansion and $\kappa_{L}$ are useful quantities. In the calculation point of view, the thermal expansion can be captured under quasi-harmonic approximation. But, the calculation of $\kappa_{L}$ requires anharmonic force constant and hence computationally expensive. A systematic study of thermal expansion and $\kappa_{L}$  are required in order to fully evaluate the TE properties of materials.

Recently, the Heusler type compounds have gained much attention in TE field. This type of compound is defined by the chemical formula $XYZ$ (half-Heuslers) or $X_{2}YZ$ (full-Heuslers), where \textit{X} and \textit{Y} are transition metals and \textit{Z} is \textit{p}-block element. These compounds are found in $L2_{1}$ phase with face-centered cubic structure. A number of experimental and theoretical studies on TE properties of these materials are given in earlier works\cite{lue_2002,lue_2004,muta,graf,sonu_jpcm,zhu,berry,shastri_1st,shastri_3rd}. Among them, Fe$_{2}$VAl is one of the full-Heusler compounds, which possesses a high power factor comparable with the state-of-the-art TE materials\cite{fva1,fva2,fva3}. Unfortunately, the \textit{ZT} is suffered by its high $\kappa$ value. In total $\kappa$, the larger contribution comes from $\kappa_{L}$ which is more than 20 W/m-K at room temperature. This high value of $\kappa_{L}$ is mainly responsible for lowering the \textit{ZT} value. In order to understand the TE properties of Fe$_{2}$VAl, the electronic structure calculations of this compound have been studied extensively. But, the phononic properties are not explored up to the scale as per our knowledge of literature survey. Therefore, a systematic phonon calculation is required to fully evaluate the TE properties of Fe$_{2}$VAl. The study of phonon related properties is expected to give significant direction for the improvement of the TE performance of Fe$_{2}$VAl. 

In this work, we have studied the thermal properties, thermal expansion and lattice thermal conductivity of Fe$_{2}$VAl through lattice dynamics calculations. Then the calculated values are compared with the available experiment. The finite displacement method with the supercell approach has been employed to calculate the phonon dispersion. The specific heat at constant volume and Helmholtz free energy are calculated under harmonic approximation. The zero-point energy and Debye temperature are calculated as $\sim$21 kJ/mol and 638 K, respectively. Thermal expansion is calculated under quasi-harmonic approximation. The room temperature value of thermal expansion is calculated as $\sim$6.3 $\times$ 10$^{-6}$ K$^{-1}$, whereas this value reaches as $\sim$7.2 $\times$ 10$^{-6}$ K$^{-1}$ at 800 K. The $\kappa_{L}$ is computed using anharmonic lattice dynamics calculation by considering the phonon-phonon interaction. At 300 K, the calculated value of $\kappa_{L}$ is $\sim$48.6 W/m-K, while it reaches $\sim$18.2 W/m-K at 800 K. The temperature dependent of phonon lifetime due to phonon-phonon interaction is also calculated. The first-principles based phonon calculations give a reasonable explanation of available experimental results.

\section{COMPUTATIONAL DETAILS}
The phonon properties were calculated using finite displacement method (FDM)\cite{fdm} and supercell approach as implemented in PHONOPY code\cite{phonopy}. A supercell of size 2 $\times$ 2 $\times$ 2 containing 128 atoms was generated by PHONOPY code. The forces on each atom of this supercell were calculated using WIEN2k\cite{wien2k} code by setting the force convergence criteria as 0.1 mRy/Bohr. A k-mesh size of 4 $\times$ 4 $\times$ 4 was used for the force calculations. These forces were used to calculate the second order force constants in order to compute the phonon dispersion using PHONOPY code. The thermal expansion was computed within quasi-harmonic approximation (QHA) as implemented in PHONOPY code.

The PHONO3PY\cite{phono3py} code was employed to calculate the lattice thermal conductivity under the supercell approach. The forces on each atom of 2 $\times$ 2 $\times$ 2 supercell were calculated using projector augmented wave (PAW) method as implemented in ABINIT software\cite{abinit} within DFT. The PAW datasets given by Jollet \textit{et. al}\cite{jollet} was used for the calculations. The local density approximation (LDA)\cite{lda} was chosen as an XC functional. A k-mesh of size 4 $\times$ 4 $\times$ 4 was used and a convergence criteria of 5 $\times$ 10$^{-8}$ Ha/Bohr was set for the self-consistent calculations. The forces obtained from ABINIT calculations were used to calculate the second and third order force constants in PHONO3PY code. These anharmonic force constants along with a dense q-mesh size of 21 $\times$ 21 $\times$ 21 were used in order to calculate lattice thermal conductivity. A real-space cutoff distance of 6.0 Bohr was set (which covers the three neighbour atoms interaction) in order to reduce the number of supercell calculations effectively.

\section{RESULTS AND DISCUSSION}                          

\subsection{PHONON PROPERTIES}
Here, we discuss the phonon dispersion, specific heat at constant volume and Helmholtz free energy of Fe$_{2}$VAl which are calculated under harmonic approximation. To check the mechanical stability of this compound, we have calculated the phonon dispersion as shown in fig. 1. The figure shows the phonon dispersion along the high symmetry direction $\Gamma-X-W-\Gamma-L$ in the first Brillouin zone. The dispersion plot consists with twelve phonon branches as the primitive cell of Fe$_{2}$VAl contains 4 atoms. All the branches give positive phonon frequency confirming the mechanical stability of the compound. Out of twelve branches, three are acoustic (plotted by red lines) and remaining nine are optical (plotted by black lines) branches. The maximum phonon energy is observed as $\sim$55 meV. Three of the optical branches having higher energy ($\sim$50 meV) are well separated from rest of the branches. At $\Gamma$-point these three optical branches are triply degenerate around $\sim$ 48.4 meV. Also on the other two points corresponding to $\sim$30.5 and 32.2 meV at $\Gamma$-point, optical branches are triply degenerate. 

The specific heat at constant volume ($C_{V}$) and Helmholtz free energy (\textit{F}) are calculated from phonon frequencies $\omega_{\textbf{q}j}$ of mode $\{\textbf{q},j\}$ using the following thermodynamic relations\cite{dove},
\begin{equation}
C_{V}={\sum\limits_{\textbf{q}j}}k_{B}\bigg(\dfrac{\hbar\omega_{\textbf{q}j}}{k_{B}T}\bigg)^{2}\frac{\exp(\hbar\omega_{\textbf{q}j}/k_{B}T)}{[\exp(\hbar\omega_{\textbf{q}j}/k_{B}T)-1]^{2}},
\end{equation} 
and
\begin{equation}
F=\dfrac{1}{2}{\sum\limits_{\textbf{q}j}}\hbar\omega_{\textbf{q}j}+k_{B}T{\sum\limits_{\textbf{q}j}}\ln[1-\exp(-\hbar\omega_{\textbf{q}j}/k_{B}T)].
\end{equation} 

Where $\hbar$ and $k_{B}$ are the reduced Planck's constant and Boltzmann constant, respectively. $T$ is the temperature. $\textbf{q}$ is wave vector and $j$ is band index. The calculated values of $C_{V}$ and \textit{F} are shown in Fig. 2 in the temperature range of $300-800$ K. The dashed line indicates the classical Dulong and Petit's limit of $C_{V}$. From the figure, it is seen that the $C_{V}$ increases rapidly with increase in temperature. But, above  $\sim$ 650 K, the $C_{V}$ is almost constant and reaches the Dulong and Petit's limit of 100 J/mol-K. The calculated $C_{V}$ is compared with the experimental data of specific heat at constant pressure ($C_{P}$) reported by Kawaharada \textit{et al.} measured by using a differential scanning calorimeter\cite{kawaharada}. From the figure, it is clear that at high temperature region, $C_{P}$ is higher than $C_{V}$ due to contribution of thermal expansion in $C_{P}$\cite{togo_2015}.  Fig. 2 also shows the temperature dependent of \textit{F}. At 0 K, the \textit{F} is found to be $\sim$21 kJ/mol and this is the zero-point energy of the compound captured under harmonic approximation in PHONOPY code. Quantum mechanically, there is a presence of vibrational energy even at zero temperature, which is called zero-point energy\cite{hao}. The Kohn-Sham DFT provides the ground state energy of the static lattice and hence ignores this zero-point energy\cite{hao}. The values of \textit{F} at 300 and 800 K are calculated as $\sim$11.5 and $\sim-$53.5 kJ/mol, respectively.

From phonon dispersion one can calculate the maximum phonon frequency. The maximum frequency is calculated as $\sim$55 meV in the $\Gamma-X$ direction of Fig. 1. This frequency is considered as Debye frequency ($\omega_{D}$) in order to calculate the Debye temperature using the formula, $\Theta_{D}$=$\hbar\omega_{D}/k_{B}$. The temperature above which all the phonon modes start to excite is called Debye temperature\cite{ashcroft}. The calculated value of $\Theta_{D}$ is $\sim$ $\sim$638 K. Therefore, above 638 K all the phonon modes are excited for Fe$_{2}$VAl. Kawaharada \textit{et al.}\cite{kawaharada} have measured the $\Theta_{D}$ experimentally from sound velocity, which is found to be $\sim$ 540 K.   

\subsection{THERMAL EXPANSION}
The definitions of thermal expansion and lattice thermal conductivity are only valid when interactions of phonon come into account. Hence, these phonon properties cann't be explained under harmonic approximation, since phonons do not interact under this approximation. Thermal expansion can be calculated by using the finite temperature lattice constants in the harmonic model, which is often known as quasi-harmonic approximation (QHA). Thermal expansion is one of the important parameters to select the materials for making TEG. Accordingly, we have calculated the linear thermal expansion coefficient ($\alpha (T)$) of Fe$_{2}$VAl under QHA. For the calculation of $\alpha (T)$, first of all the total free energy as a function of primitive cell volume has been calculated at different temperatures as shown in Fig. 3(a). Here, the total free energy is expressed as: $F(T; V) = [U_{el}(V)-U_{el}(V_{0})]+F_{ph}(T; V)$ at given temperature and volume. Where $U_{el}(V)-U_{el}(V_{0})$ is the relative ground state electronic energy, $V_{0}$ is the equilibrium volume at 0 K. $F_{ph}(T; V)$ is the phonon Helmholtz free energy. From Fig. 3(a), it is observed that every plot corresponding to each temperature gives the energy minima (denoted by solid green square). These energy minima are signifying the equilibrium volumes of primitive cell at different temperatures which are connected by solid line (red). Then, these equilibrium volumes are plotted as a function of temperature as shown in Fig. 3(b). Figure shows that in the lower temperature range of $\sim0-100$ K, the primitive cell volume is almost constant. But, above 100 K, the volume is increasing linearly with rising in temperature. The volumes of primitive cell are calculated as $\sim$45.1 and $\sim$45.7 \AA$^{3}$ at 0 and 800 K, respectively.

\begin{figure}
\includegraphics[width=0.8\linewidth, height=5.8cm]{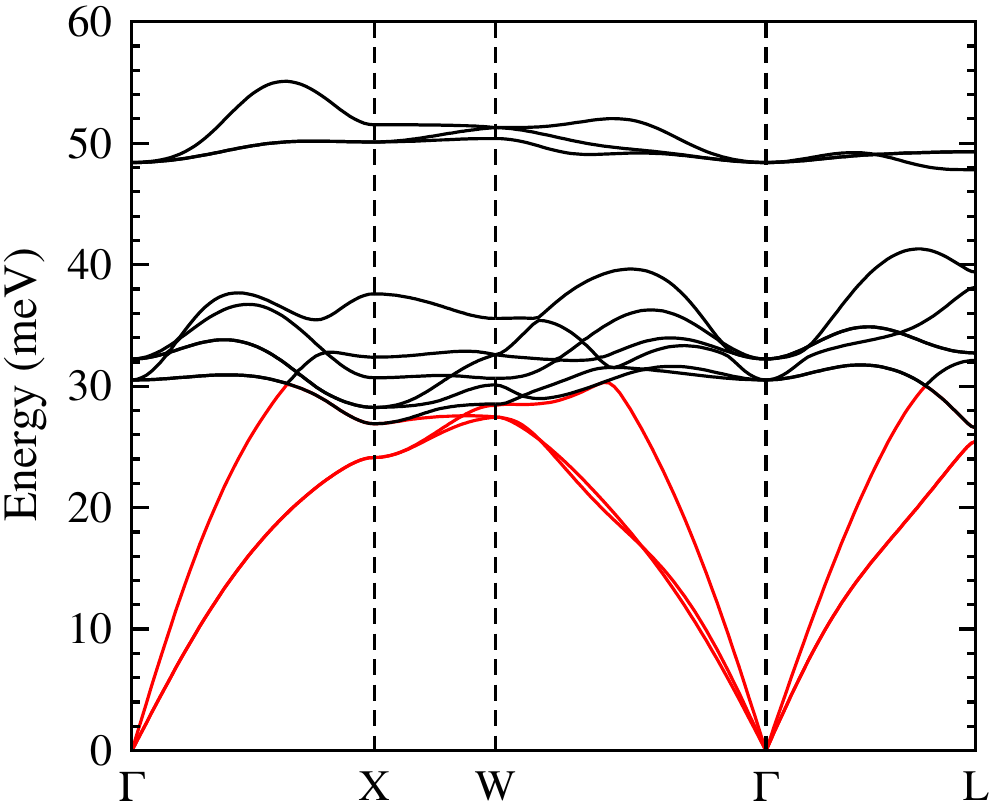} 
\caption{\small{Calculated phonon dispersion.}}
\end{figure}

\begin{figure}
\includegraphics[width=0.8\linewidth, height=6.0cm]{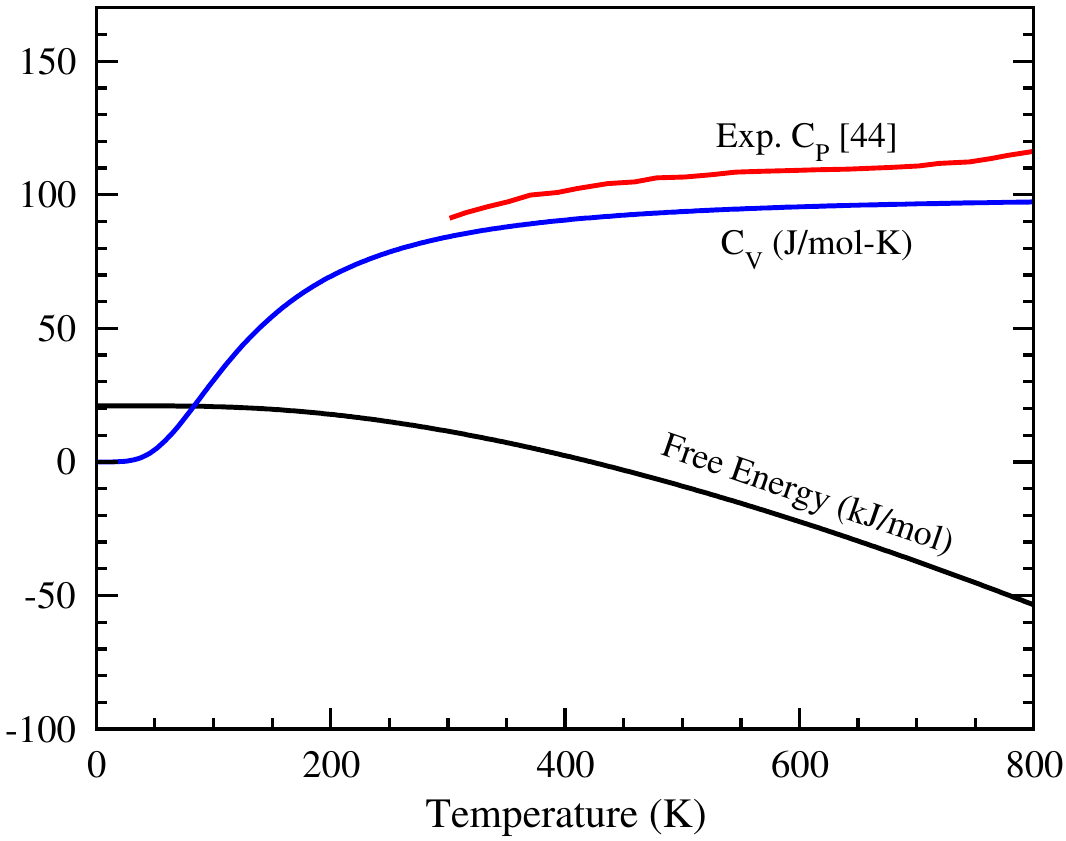} 
\caption{\small{Specific heat at constant volume ($C_{V}$) and Helmholtz free energy ($F$) along with experimental specific heat at constant pressure ($C_{P}$)\cite{kawaharada}.}}
\end{figure}

\begin{figure}
\includegraphics[width=0.95\linewidth, height=5.5cm]{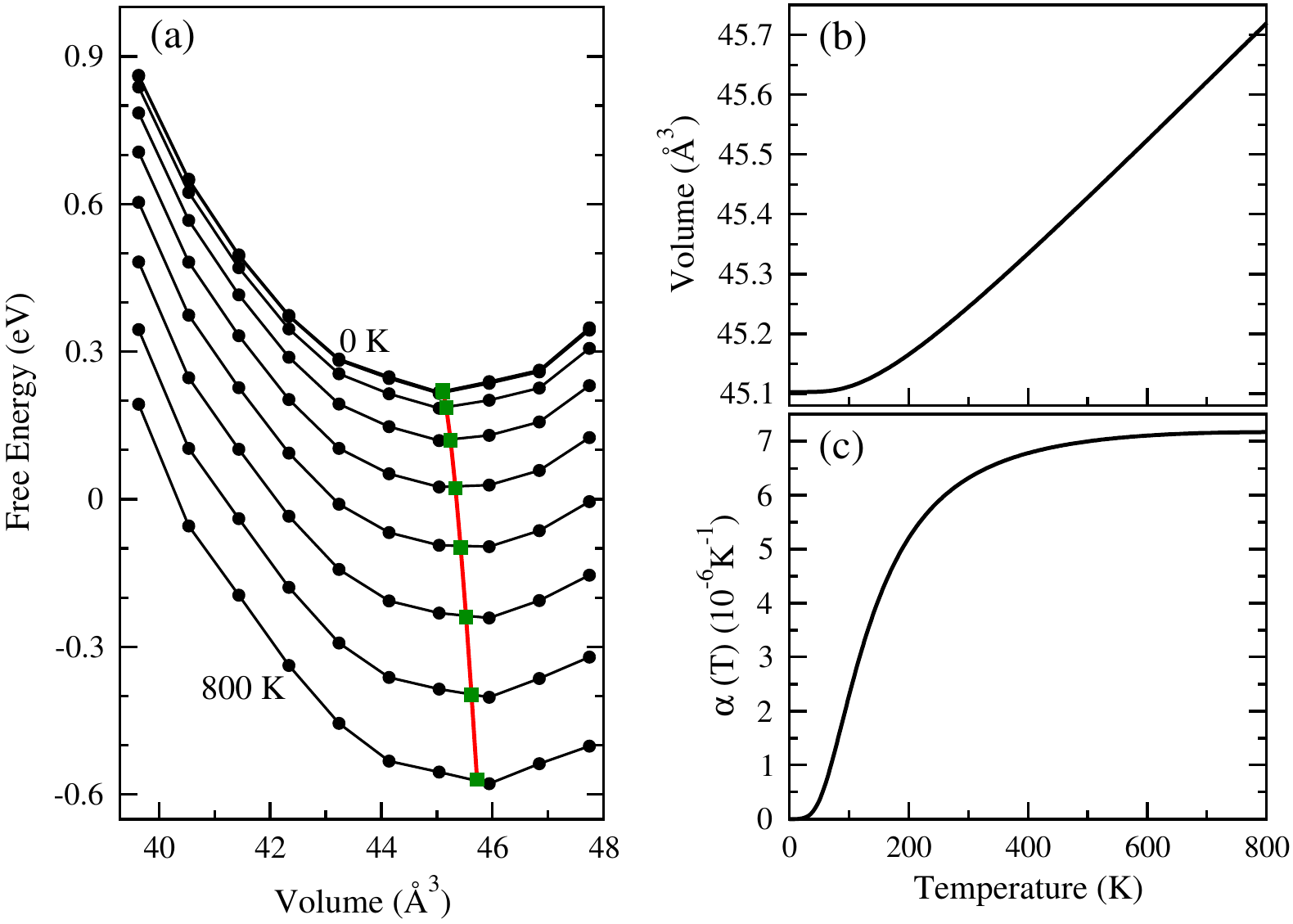} 
\caption{\small{(a) Variation of total free energy F with primitive cell volume. (b) Change in primitive cell volume with temperature. (c) Linear thermal expansion coefficient $\alpha (T)$ as a function of temperature.}}
\end{figure}

The volumetric thermal expansion coefficient is calculated using the formula: $\beta(T) = \frac{1}{V(T)}\frac{\partial V(T)}{\partial T}$. Where, $V(T)$ is the volume of a primitive cell as a function of temperature which is taken from Fig. 3(b). Fe$_{2}$VAl possesses cubic structure. Therefore, by considering the equal expansion in all the three directions, the $\alpha (T)$ can be calculated as one third of $\beta(T)$\cite{ashcroft}. Fig. 3(c) shows the calculated $\alpha (T)$ in the temperature range $0-800$ K. The values of $\alpha (T)$ increase rapidly up to $\sim$300 K, then it's increment rate decreases. At high temperature region (above $\sim$ 500 K), the values of $\alpha (T)$ are almost constant. The room temperature value of $\alpha (T)$ is calculated as $\sim$6.3 $\times$ 10$^{-6}$ K$^{-1}$, whereas this value reaches as $\sim$7.2 $\times$ 10$^{-6}$ K$^{-1}$ at 800 K. We could not go through any work of experimental thermal expansion of Fe$_{2}$VAl in the studied temperature range. Therefore, to verify our theoretical calculation of $\alpha (T)$ under QHA, an experimental measurement is required. In the application point of view, the study of $\alpha (T)$ is much needful before putting the material for designing the TEG. Because, thermal expansion is responsible for cracking and porosity of the materials and hence affect the performance of TEG\cite{zhang}.  


\subsection{LATTICE THERMAL CONDUCTIVITY}
The lattice part of thermal conductivity ($\kappa_{L}$) of Fe$_{2}$VAl is calculated using the linearized phonon Boltzmann equation (LBTE) under single-mode relaxation time (SMRT) method as implemented in PHONO3PY code\cite{phono3py}. The forces on each atom are computed using first-principle calculation. Then, anharmonic force constants are calculated using PHONO3PY in order to calculate the imaginary part of phonon self-energy. Here, it is important to note that only the phonon-phonon interactions are considered in the calculation. The phonon lifetime of mode $\lambda$ is obtained from the imaginary part of phonon self-energy as follows\cite{maradudin,togo_2015}:    
\begin{equation}
\tau_{\lambda}=\frac{1}{2\Gamma_{\lambda}(\omega_{\lambda})},
\end{equation}
where, $2\Gamma_{\lambda}(\omega_{\lambda})$ is the phonon linewidth of mode $\lambda$ and $\omega_{\lambda}$ is the frequency of phonon of mode $\lambda$. The computing of this $\tau_{\lambda}$ is the more challenging in order to calculate the $\kappa_{L}$. Under the approximation of SMRT, the $\kappa_{L}$ is calculated using the following LBTE\cite{togo_2015,srivastava}: 
\begin{equation}
\kappa_{L}=\frac{1}{NV_{0}}\sum_{\lambda}C_{\lambda}\textbf{v}_{\lambda}\otimes \textbf{v}_{\lambda}\tau_{\lambda}^{SMRT},
\end{equation}
where, \textit{N} and $V_{0}$ are the number of unit cells and volume of a unit cell, respectively. $\textbf{v}_{\lambda}$ and $C_{\lambda}$ are the group velocity and heat capacity of phonon of mode $\lambda$. $\tau_{\lambda}^{SMRT}$ is the single-mode relaxation time of phonon mode $\lambda$. This $\tau_{\lambda}^{SMRT}$ is taken as the phonon lifetime $\tau_{\lambda}$ for calculating $\kappa_{L}$, where $\tau_{\lambda}$ is defined in Eqn. 3.     

Fig. 4 shows the calculated values of $\kappa_{L}$ along with the experimental data for Fe$_{2}$VAl in the temperature range $300-800$ K. The experimental $\kappa_{L}$ is obtained as: $\kappa_{L} = \kappa - \kappa_{e}$. Where, $\kappa_{e}$ is calculated from Wiedeman-Franz law: $\kappa_{e}=L\sigma T$, \textit{L} is Lorenz number. The experimental $\sigma$ and $\kappa$ are taken from the work of Sk \textit{et el}\cite{shamim_fva}. At 300 K, the calculated value of $\kappa_{L}$ is $\sim$48.6 W/m-K, while it reaches $\sim$18.2 W/m-K at 800 K. The corresponding experimental values of $\kappa_{L}$ are $\sim$22.8 W/m-K and $\sim$11.0 W/m-K at 300 and 800 K, respectively. The values of calculated and experimental $\kappa_{L}$ come closer as the temperature increases. But, at a lower temperature region, a reasonably high deviation between calculated and experimental $\kappa_{L}$ is observed from the figure. This deviation can be attributed by many factors. For instance, it is often seen that the synthesized polycrystalline sample contains defects and/or disorder. In addition to this, the grain size, grain boundary and sample preparation condition are the major factors which affect the thermal conductivity of sample. In contrast, the calculation of $\kappa_{L}$ has been done on single crystalline Fe$_{2}$VAl, where only the phonon-phonon interactions are considered. But, in the real sample, relaxation time of phonon also depends on phonon-electron interactions, phonon-defect interactions etc. In addition to this, the ground state phonon dispersion is used to calculate the temperature dependent $\kappa_{L}$. But, in reality the phonon dispersion is temperature dependent. These all are the possible reasons for the deviation between calculated and experimental values of $\kappa_{L}$. Considering all the interactions present in a real sample and temperature dependent phonon dispersion may improve the calculated $\kappa_{L}$. But, such calculations require sufficiently more computational cost and also beyond the scope of our present study. 

One of the key quantities for understanding the phononic transport properties is lifetime of phonon. The lifetime of phonon depends on various scattering mechanisms, \textit{viz.} phonon-phonon interaction (PPI), electron-phonon interaction, phonon-defect interaction etc. Here, we have calculated the phonon lifetime due to PPI in order to understand the feature of $\kappa_{L}$ in the temperature range of $300-800$ K. The phonon lifetime of mode $\lambda$ is calculated from the imaginary part of phonon self-energy using eq. 3. Under the many-body perturbation theory, the imaginary part of phonon self-energy can be computed by the formula\cite{togo_2015},         
\begin{eqnarray}
\Gamma_{\lambda}(\omega)=\frac{18\pi}{\hbar^{2}}\sum_{\lambda^{'}\lambda^{''}}|\Phi_{-\lambda\lambda^{'}\lambda^{''}}|^{2}\{(n_{\lambda^{'}}+n_{\lambda^{''}}+1) \nonumber \\
\times \delta(\omega-\omega_{\lambda^{'}}-\omega_{\lambda^{''}})+(n_{\lambda^{'}}-n_{\lambda^{''}}) \nonumber \\
\times [\delta(\omega+\omega_{\lambda^{'}}-\omega_{\lambda^{''}})-\delta(\omega-\omega_{\lambda^{'}}-\omega_{\lambda^{''}})]\},
\end{eqnarray}  
where $\Phi_{-\lambda\lambda^{'}\lambda^{''}}$ is the strength of interaction among three phonons $\lambda$, $\lambda^{'}$ and $\lambda^{''}$ involved in the scattering. $\omega_{\lambda}$ is the frequency of phonon mode $\lambda$. $n_{\lambda}$ is the phonon occupation number at the equilibrium,
\begin{equation}
n_{\lambda}=\frac{1}{\exp(\hbar \omega_{\lambda}/k_{B}T)-1}.
\end{equation}

Fig. 5(a) displays the lifetime of different phonon branches marked as b1, b2, b3, ..., b12. Among them, b1, b2 and b3 are acoustic branches and the remaining are optical branches. The lifetime is calculated by taking the weight average over \textit{q}-point in the temperature range of $300-800$ K. The branch having higher lifetime will have the lesser scattering. From the figure, it is clear that the optical branch b7 has the highest lifetime indicating the lowest scattering. Three higher energy optical branches b10, b11 and b12 show the lower lifetime as compared to other branches signifying the higher scattering. The lifetime of all the branches decreases as the temperature increases. This suggests that at high temperature region a phonon faces more scattering with other phonons as compared to low temperature region. We have also calculated the lifetime of total, acoustic and optical branches by taking the average of corresponding number of phonon branches as shown in Fig 5(b). The acoustic branches show the higher lifetime as compared to optical branches. But, as the temperature increases, both the lifetimes of acoustic and optical branches come closer. The total phonon lifetimes are calculated as $\sim$9.57$\times$10$^{-12}$ and $\sim$3.37$\times$10$^{-12}$ s at 300 and 800 K, respectively. For calculating $\kappa_{L}$ (eq. 4), the group velocity is taken as temperature independent. Fig. 4 shows that the specific heat increases as the temperature increases and saturates at high temperature (above $\sim$600 K). Therefore, the decrement behaviour of $\kappa_{L}$ (Fig. 4) is directly related with the decreasing trend of phonon lifetime with temperature.

\begin{figure}
\includegraphics[width=0.8\linewidth, height=5.9cm]{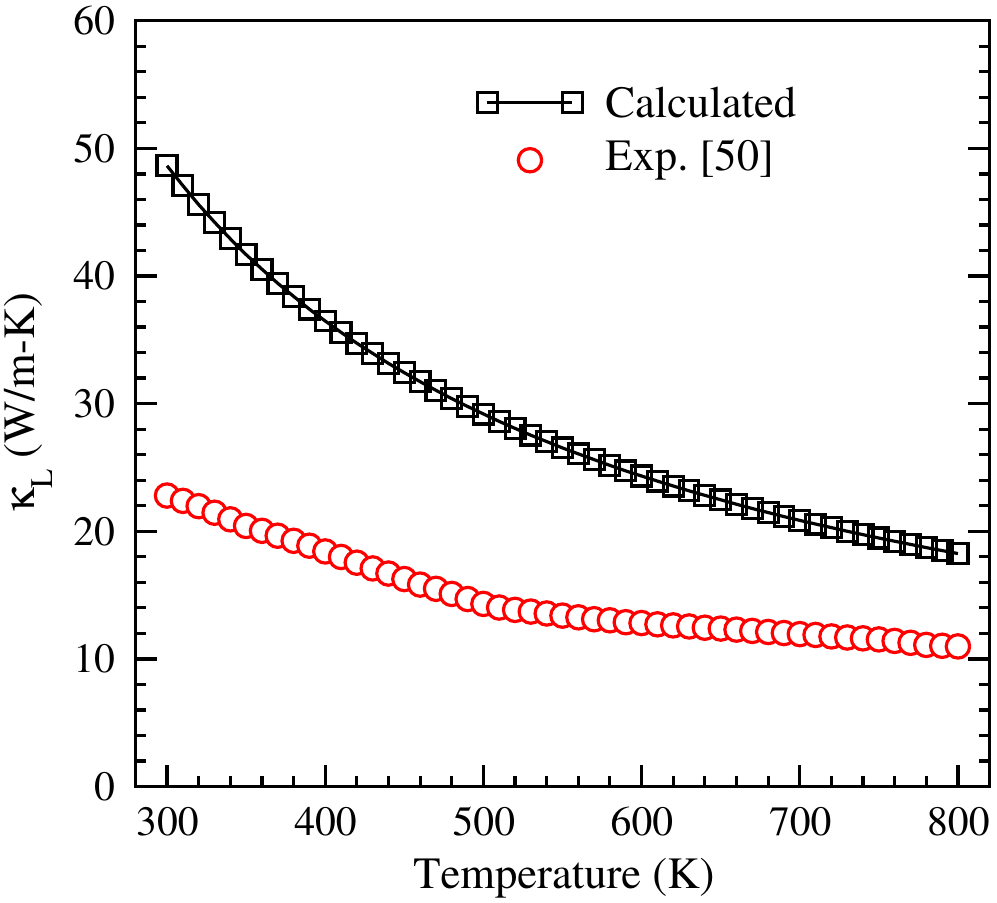} 
\caption{\small{Comparison of calculated and experimental\cite{shamim_fva} lattice thermal conductivity as a function of temperature.}}
\end{figure}

\begin{figure}
\includegraphics[width=0.8\linewidth, height=11.0cm]{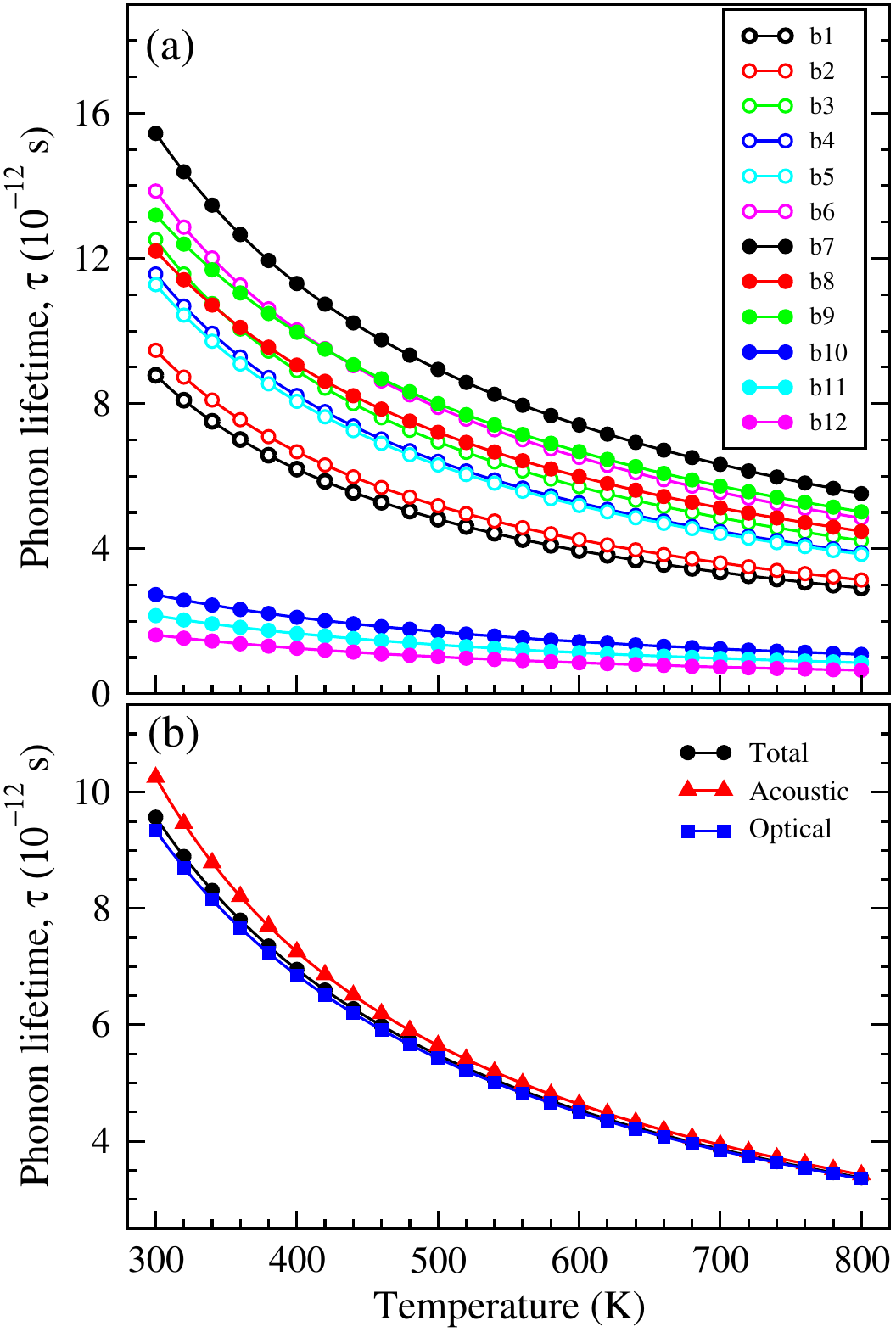} 
\caption{\small{(a) The phonon lifetime for twelve phonon branches. (b) Total, acoustic, optical phonon lifetime due to PPI as a function of temperature.}}
\end{figure}

\section{CONCLUSIONS}
In conclusion, the phonon related properties of Fe$_{2}$VAl are calculated using DFT based methods and compared with the experiment. The phonon dispersion, specific heat at constant volume and Helmholtz free energy of Fe$_{2}$VAl are calculated under harmonic approximation. The phonon dispersion shows the positive phonon frequencies of all the branches signifying the mechanical stability of the compound. The zero-point energy and Debye temperature are computed as $\sim$21 kJ/mol and 638 K, respectively. Under the quasi-harmonic approximation, the thermal expansion ($\alpha (T)$) is calculated, while lattice thermal conductivity ($\kappa_{L}$) is computed using first-principle anharmonic lattice dynamics calculations. The values of $\alpha (T)$ increase rapidly up to $\sim$300 K ($\sim$6.3 $\times$ 10$^{-6}$ K$^{-1}$), then increase slowly. Above $\sim$ 500 K, the values of $\alpha (T)$ are almost constant with the value of $\sim$7.2 $\times$ 10$^{-6}$ K$^{-1}$ at 800 K. At 300 K, the calculated ($\sim$48.6 W/m-K) $\kappa_{L}$ are found to deviate from the experiment ($\sim$22.8 W/m-K). As the temperature increases the calculated and experimental values of $\kappa_{L}$ come closer to each other with the respective values of $\sim$18.2 W/m-K and $\sim$11.0 W/m-K at 800. The temperature dependent trend of $\kappa_{L}$ is understood through the calculation of phonon lifetime due to phonon-phonon interaction. This study proposes that DFT based lattice dynamics calculations can be used to address the experimental phonon properties of materials at high temperature.

\end{document}